\documentclass[letter]{aa} 
\usepackage{hyperref}
\usepackage{lscape}
\usepackage{xcolor}
\definecolor{dark-red}{rgb}{0.9,0.0,0.0}
\definecolor{dark-blue}{rgb}{0.15,0.15,0.5}
\definecolor{dark-green}{rgb}{0.15,0.4,0.15}
\definecolor{medium-blue}{rgb}{0,0,0.9}
\hypersetup{
    colorlinks, linkcolor=medium-blue,
    citecolor={dark-blue} , urlcolor={dark-green}
}
\usepackage{changepage} 
\usepackage{graphicx}
\usepackage{url}
\usepackage{upgreek}

\makeatletter
\renewcommand*\aa@pageof{, page \thepage{} of \pageref*{LastPage}}
\makeatother

\usepackage{txfonts}
\begin{document}

   \title{Radial Velocity Survey for Planets around Young stars (RVSPY)}

   \subtitle{A transiting warm super-Jovian planet around HD\,114082,\\a young star with a debris disk\thanks{Based on observations collected at the European Organization for Astronomical Research in the Southern Hemisphere under MPG programs 0101.A-9012, 0103.A-9010, 0104.A-9003, 0107.A-9004, 0108.A-9014, 0109.A-9014 and ESO programs 098.C-0739, 1101.C-0557.}
  }

   \author{Olga V. Zakhozhay\inst{1,2}
          \and
          Ralf Launhardt\inst{1}
          \and
          Trifon Trifonov\inst{1,3,4}
           \and
          Martin K\"urster\inst{1}
               \and
          Sabine Reffert\inst{4}
           \and
          Thomas Henning\inst{1}
            \and
         Rafael Brahm\inst{5,6,7}
           \and 
          José I. Vinés\inst{8}
           \and 
         Gabriel-Dominique Marleau\inst{9,10,11,1}
            \and
           Jayshil A. Patel\inst{12}
          }
          
 \institute{Max-Planck-Institut f\"{u}r Astronomie,\ K\"{o}nigstuhl  17, 69117 Heidelberg, Germany\\ 
              \email{zakhozhay@mpia.de}
         \and
          Main Astronomical Observatory, National Academy of Sciences of the Ukraine, 03143 Kyiv, Ukraine
          \and     
          Department of Astronomy, Sofia University ``St Kliment Ohridski'', 5 James Bourchier Blvd, BG-1164 Sofia, Bulgaria
          \and     
          Landessternwarte, Zentrum f\"{u}r Astronomie der Universit\"{a}t Heidelberg,\ K\"{o}nigstuhl 12, 69117 Heidelberg, Germany
          \and      
          Facultad de Ingeniera y Ciencias, Universidad Adolfo Ib\'{a}\~{n}ez, Av. Diagonal las Torres 2640, Pe\~{n}alol\'{e}n, Santiago, Chile
          \and
          Millennium Institute of Astrophysics (MAS), Nuncio Monseñor Sótero Sanz 100, Providencia, Santiago, Chile
          \and
          Data Observatory Foundation, DO; Diagonal Las Torres Nº2460, Building E, Peñalolén, Santiago, Chile
          \and
          Departamento de Astronom\'ia, Universidad de Chile, Casilla 36-D, Santiago, Chile
          \and
          Fakult\"at f\"ur Physik, Universit\"at Duisburg-Essen, Lotharstraße 1, 47057 Duisburg, Germany
          \and
         Institut f\"{u}r Astronomie und Astrophysik, Universit\"{a}t T\"{u}bingen, Auf der Morgenstelle 10, 72076 T\"{u}bingen, Germany
          \and
         Physikalisches Institut, Universit\"{a}t Bern, Gesellschaftsstr. 6, 3012 Bern, Switzerland
        \and
        Institutionen f\"or astronomi, AlbaNova universitetscentrum, Stockholms universitet, SE-106 91 Stockholm, Sweden
          }
          

   \date{Received August 08, 2022; accepted October 19, 2022}

 
  \abstract
   {}
   {We aim to detect planetary companions to young stars with debris disks via the radial velocity method.%
   }
   {We observed HD\,114082 during April 2018 - August 2022 as one of the targets of our RVSPY program (Radial Velocity Survey for Planets around Young stars). We use the FEROS spectrograph, mounted to the MPG/ESO 2.2\,m telescope in Chile, to obtain high signal-to-noise spectra and time series of precise radial velocities (RVs). Additionally, we analyze archival HARPS spectra and TESS photometric data. We use the CERES, CERES++ and SERVAL pipelines to derive RVs and activity indicators and ExoStriker for the independent and combined analysis of the RVs and TESS photometry.}
   {We report the discovery of a warm super-Jovian companion around HD\,114082 based on a 109.8$\pm$0.4\,day signal in the combined RV data from FEROS and HARPS, and on one transit event in the TESS photometry. The best-fit model indicates a 8.0$\pm 1.0$\,${\rm M_{Jup}}$ companion with a radius of 1.00$\pm 0.03{\rm R_{Jup}}$ in an orbit with a semi-major axis of 0.51$\pm$0.01\,au and an eccentricity of 0.4$\pm$0.04. The companions orbit is in agreement with the known near edge-on debris disk located at $\sim$28\,au.
   HD\,114082\,b is possibly the youngest (15$\pm$6\,Myr), and one of only three young (<100\,Myr) giant planetary companions for which both their mass and radius have been determined observationally. It is probably the first properly model-constraining giant planet that allows distinguishing between hot and cold-start models. It is significantly more compatible with the cold-start model.
     }
   {}


   \maketitle
%

\section{Introduction}
\label{sec:Intro}

The search for exoplanets is one of the fastest-developing topics of modern astronomy: We learn more details about newly discovered planets, their properties, and their evolution nearly every day. More than five thousand planets were detected by mid-2022 (see exoplanet.eu; \citealt{schneider2011}). Only for $\sim$10\% of these have the masses and radii been determined directly through observations, however.
Only two systems, V\,1298\,Tau \citep{SuarezMascareno2022} and AU\,Mic \citep{plavchan2020,martioli2021,zicher2022}, are younger than 50\,Myr and have known planetary companions with such determinations. To constrain evolutionary models, we therefore need to know both quantities not only for a larger number of planets, but also for planets that are significantly younger ($\ll$100\,Myr) than those for which we currently have both parameters.

In 2017, we started a large radial velocity (RV) survey for planets around young stars (RVSPY) with debris disks using the FEROS spectrograph at the 2.2\,m telescope at the La Silla Observatory in Chile \citep{zakhozhay2022}.  This survey aims to supplement the large direct Imaging Survey for Planets around Young stars (NaCo-I\,SPY), which was carried out between 2015 and 2019  \mbox{\citep{launhardt2020}}. The two surveys overlap for 54 targets. In addition to our own observational data, we also use complementary data from HARPS\footnote{HARPS is the High Accuracy Radial velocity Planet Searcher, mounted at the 3.6\,m telescope at ESO's La Silla observatory in Chile \mbox{\citep{Mayor2003}}.}, TESS\footnote{TESS is the Transiting Exoplanet Survey Satellite \mbox{\citep{Ricker2015}}.} , and GAIA\footnote{GAIA is the Global Astrometric Interferometer for Astrophysics \mbox{\citep{gaia_mission}}.}.

In this paper, we report the discovery of a super-Jovian companion to HD\,114082, a young (15$\pm$6\,Myr, \citealt{pm2016}) F3V star in the lower Centaurus Crux association 
\citep{gagne2018} (see Table\,\ref{tab:phys_pars}). The star is surrounded by a spatially resolved debris disk that has the form of a narrow dust ring at $r\sim$28\,au and an inclination with respect to the line of sight of 83.3$^{+0.4}_{-3.8}$\,degree \citep[SPHERE;][]{wahhaj2016}.
HD\,114082 was also observed in the RVSPY high-cadence survey, but no hot Jupiter with $P<10$\,d was detected down to a 3\,$\sigma$\ detection limit of 2-3\,M$_{\rm Jup}$\ \citep{zakhozhay2022}. In addition, it was also observed in the $L^{\prime}$-band high-contrast direct-imaging I\,SPY survey, but no wide-orbit companion was detected down to 3\,$\sigma$\ detection limits of 40\,M$_{\rm Jup}$\ at 10\,au to 4\,M$_{\rm Jup}$\ at 100\,au and beyond \citep{launhardt2020}.

 \begin{table}
 \caption{Basic stellar parameters of HD\,114082}
\label{tab:phys_pars}
\centering          
\begin{tabular}{llll}
\hline\hline
Parameter & Unit & Value & Ref.\\
\hline
Distance                      & [pc]                & 95.06$\pm$0.20     & 1 \\
$V$\                       & [mag]               & 8.21$\pm$0.01      & 2 \\
SpT                        & \ldots                 & F3\,V              & 3 \\
$T_{\rm eff}^{\rm sp}$     & [K]                 & 6651$\pm$35\tablefootmark{(1)}        & 5 \\
$T_{\rm eff}^{\rm ph}$     & [K]                 & 6600$\pm$70        & 4 \\
$[{\rm Fe/H}]$             & \ldots               & 0.00$\pm$0.03\tablefootmark{(1)}       & 5 \\
$\log(g)$                  & $\log$[cm\,s$^{-2}$] & 4.19$\pm$0.03\tablefootmark{(1)}       & 5 \\
$M_{\ast}$                 & [M$_{\odot}$]       & 1.47$\pm$0.07      & 4 \\
$L_{\ast}$                 & [L$_{\odot}$]       & 3.83$\pm$0.05      & 4 \\
$R_{\ast}$                 & [R$_{\odot}$]       & 1.49$\pm$0.05      & 4 \\
$v\sin(i)$                 & [km\,s$^{-1}$]      & 39.2$\pm$0.5\tablefootmark{(1)}        & 5 \\
$P_{\rm rot}^{\rm max}$    & [d]                 & 1.924$\pm$0.016\tablefootmark{(2)}     & 5  \\
Assoc.                     & \ldots                 & Lower Centaurus Crux                & 4, 6 \\
Age                        & [Myr]               & 15$\pm$6\tablefootmark{(3)}            & 4, 7 \\

$R_{\rm disk}$          & [au]                & 27.7$\pm$3         & 8 \\
$i_{\rm disk}$        & [deg]               & 83.3$\pm$3         & 8 \\
\hline 
\end{tabular}

\tablefoot{
\tablefoottext{1}{he value derived using the \texttt{ZASPE} pipeline~\citep{Brahm2017b} on the observed spectra.}
\tablefoottext{2}{There is no unique $P_{\rm rot}$\ visible in the TESS photometry. $P_{\rm rot}^{\rm max}$\ is calculated from $v\sin(i)$\ assuming $i=90\degr$. $P_{\rm rot}$\ should be 1.91\,d to match $i_{\rm disk}$.}
\tablefoottext{3}{As uncertainty,
we adopt the age spread \citet{pm2016} derived for the lower Centaurus Crux association rather than the uncertainty of the mean association age (3\,Myr).}
}
\tablebib{
(1)~\citet{gaia_mission,eGDR3}; 
(2)~\citet{Tycho2};
(3)~Simbad;
(4)~\citet{zakhozhay2022};
(5)~This paper;
(6)~\citet{gagne2018};
(7)~\citet{pm2016};
(8)~\citet{wahhaj2016}.
}
\end{table}


\section{Observations and data reduction}
\label{sec:Obs}

HD\,114082 was observed with FEROS \citep{Kaufer99}, a high-resolution, environmentally controlled Échelle spectrograph that is mounted at the 2.2\,m telescope at La Silla Observatory in Chile. This instrument has a high spectral resolution (R=48\,000), high efficiency ($\sim$20\%), and large wavelength coverage (350\,--\,920\,nm). The observations were carried out in the object-calibration mode, with one fiber on the star and the other fiber fed by a ThAr+Ne calibration lamp. The observations were carried out between April 17, 2018, and August 20, 2022. During this period, we obtained seven sets of high-cadence time series\footnote{Each set consists of 5--20 observations taken in consecutive nights whenever possible.} with 63 useful spectra in total. %
We  did not include 3 archival\footnote{\url{http://archive.eso.org/eso/eso_archive_main.html}} spectra taken on the single night of May 30, 2009, that is,\ almost nine years before the RVSPY survey. These observations were obtained with a different calibration plan. \looseness=-5

The spectra were reduced and the RVs and bisector spans (BS)\footnote{The bisector span is a measure of the asymmetry of absorption lines \citep{Queloz01}.} were derived in a semi-automatic fashion with the \texttt{CERES} pipeline \citep{Brahm2017a}. The stellar activity indicators (full width at half maximum (FWHM), H$_\alpha$, HeI, and NaI\,D$_1$+D$_2$ indices) were derived from the FEROS spectra using the \texttt{CERES-plusplus} \texttt{Python} package{\footnote{\url{https://github.com/jvines/Ceres-plusplus}}}.\looseness=-5

In addition, we downloaded 18 HARPS spectra that are publicly available in the ESO archive (at the time of writing). We obtained precise RVs and stellar activity indicators from the \texttt{HARPS-RVBank}\footnote{\url{www.mpia.de/homes/trifonov/HARPS_RVBank.html}} \citep{Trifonov2020}, which was updated for this purpose, to include the latest spectra that became public after 2020. All data (RV, H$_\alpha$, NaI\,D, NaII\,D, differential line width dLW, and chromatic RV index CRX\footnote{Chromatic index is RV gradient as a function of wavelength with the RVs measured in the echelle orders of the spectrograph \citep{Zechmeister2018}.}) were computed using the \texttt{SERVAL} pipeline \citep{Zechmeister2018}.

Furthermore, we found that HD\,114082 is scheduled for TESS observations in four sectors: sectors 11, 38, 64, and 65\footnote{\url{https://heasarc.gsfc.nasa.gov/cgi-bin/tess/webtess/wtv.py}}. The data for the first two sectors are already public, but only for sector 38 the light curves are available in the MAST database\footnote{\url{https://archive.stsci.edu/missions-and-data/tess}}. They have a cadence of 2~minutes. We tried to reduce sector 11 using the available raw data, but were not able to extract a light curve. The reason most probably is that the star lies close to the edge of the CCD in this sector. To verify that the sector\,38 TESS light curve is not blended by another (known) star, we inspected the TESS target pixel file image \citep[TPF;][]{Aller2020}  of sector\,38 and used the automatic tool \texttt{tpfplotter}\footnote{\url{https:www.github.com/jlillo/tpfplotter}} to overlay the Gaia EDR3 catalog on the TESS TPF (see Fig.\,\ref{fig:TPF_S38}). 
The observations in sectors 64 and 65 will be carried out only in April and May 2023, respectively.


 \section{Results}
  \label{sec:res}

\begin{figure}[!t]  
\includegraphics[width=0.49\textwidth]{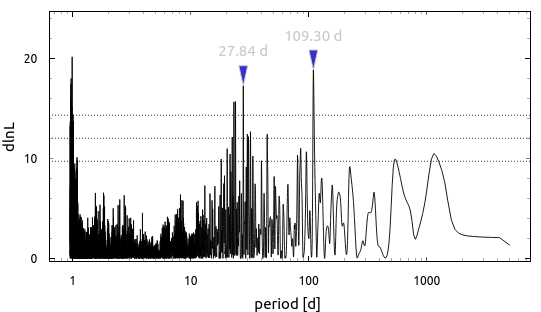}
\centering
\caption{\label{fig:RVgls}
MLP power of the combined RV data (FEROS and HARPS) for HD\,114082. Horizontal lines reflect the 0.1\%, 1\%, and 10\% false-alarm probabilities (from top to bottom).
}
\end{figure}

We inspected the RV time series of both the FEROS and the HARPS data and performed a combined analysis using the exoplanet toolbox \texttt{Exo-Striker}\footnote{\url{https://github.com/3fon3fonov/exostriker}}~\citep{Trifonov2019}. For the period search, we computed the maximum likelihood periodogram \citep[MLP;][]{Baluev2008,Baluev2009}, which is equal to fitting sine waves. This allows the determination of RV jitter for both instruments as well as the RV offset between the instruments. Visual examination of the data indicates evidence for a linear trend. We therefore computed the MLP for the detrended data.  Fig.\,\ref{fig:RVgls} shows the resulting MLP power spectrum. Four significant peaks are seen around 23\,d, 24\,d, 28\,d, and 109\,d. These peaks  mostly come from the FEROS  data sample, which is more numerous and has a larger baseline. The 28 d signal is close to the fourth harmonic of the most significant peak at 109\,d and possibly caused by an eccentric orbit. %
The peaks near 1\,d are related to the data sampling and are daily aliases of the mentioned significant signals. %
The two peaks at 23 and 24\,d cannot be explained by aliases and are explored in Sect.\,\ref{ssec:res:anacomb}.

The sector 38 TESS light curve includes a single transit event, suggesting the presence of an $\approx$1$R_{\rm Jup}$-sized companion with a mass between $0.3\,{\rm M}_{\rm Jup}$\ and the brown dwarf mass regime, and with a minimum orbital period of $P^{\prime}_{\rm min} = 190_{-140}^{+100}$\,d. Details of these analyses and estimates are described in Sect.\,\ref{ssec:app:anasep}.
Both the orbital period range and the mass range of the transiting companion overlap with the corresponding values inferred directly from modeling the RV data alone, which yields $P$ = 109.3\,d and $M_{\rm pl}=8.6\,{\rm M}_{\rm Jup}$\ (model not shown here). Consequently, we have reason to assume that both signals are caused by the same companion. We therefore attempt a combined analysis of both data sets.


 \subsection{Combined analysis of RV data and TESS photometry}
 \label{ssec:res:anacomb}

For the joint-fit analysis of transit and RV data, we used \texttt{Exo-Striker} , adopted a Keplerian model, and allowed a linear trend in the RV data. The fit parameters for the planet are orbital period $P$, eccentricity $e$, semi-amplitude $K$, argument of periastron $\omega$, and the mid-transit time $t_0$. %
Additional free parameters were RV and transit photometry data offsets (RV$_{\rm off,FEROS}$, RV$_{\rm off,HARPS}$, and transit$_{\rm off,TESS}$), jitter values (RV$_{\rm jitt,FEROS}$, RV$_{\rm jitt,HARPS}$ , and transit $_{\rm jitt,TESS}$), and the quadratic limb-darkening coefficients ($u_1$ and $u_2$). Details of the fitting procedure and the parameter priors can be found in Appendix~\ref{ssec:app:mod}.

\begin{figure*}[htb]
\includegraphics[width=0.95\textwidth]{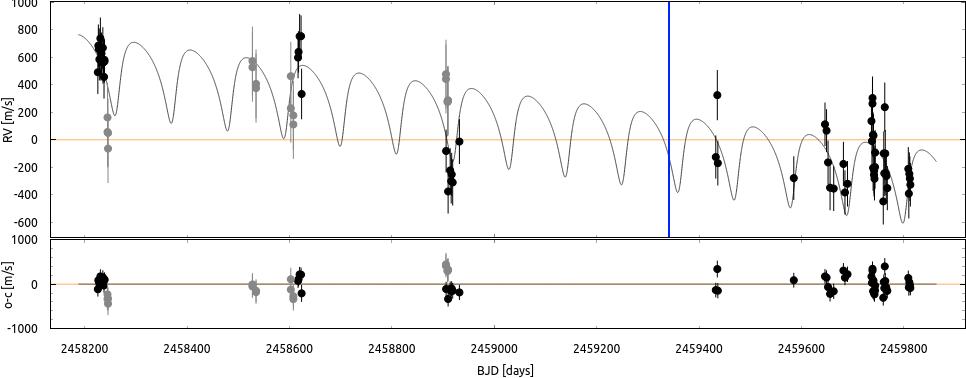}
\centering
\caption{\label{fig:RV}
RV time series data for HD\,114082. FEROS data from RVSPY are shown as black circles. The RVs computed from the publicly available HARPS spectra are shown as gray circles. Error bars account for the suspected activity jitter (see text for details). The solid gray line shows the best-fit model (see parameters in Table\,\ref{tab:plan_pars}), and the vertical blue line indicates the time of the observed transit. The lower panel shows the residuals of the fit.
}
\end{figure*}

\begin{figure}[htb]
\includegraphics[width=0.4\textwidth]{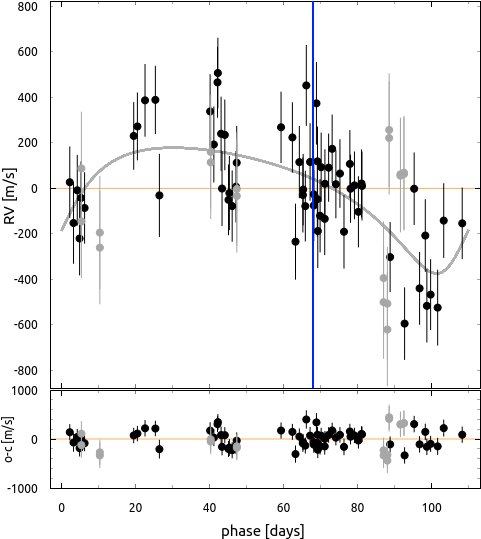}
\centering
\caption{\label{fig:phase_pl1}
Phase-folded version of Fig.\,\ref{fig:RV}, folded with the period of planet b, 109.8\,days. 
}
\end{figure}

\begin{figure}[htb]
\includegraphics[width=0.44\textwidth]{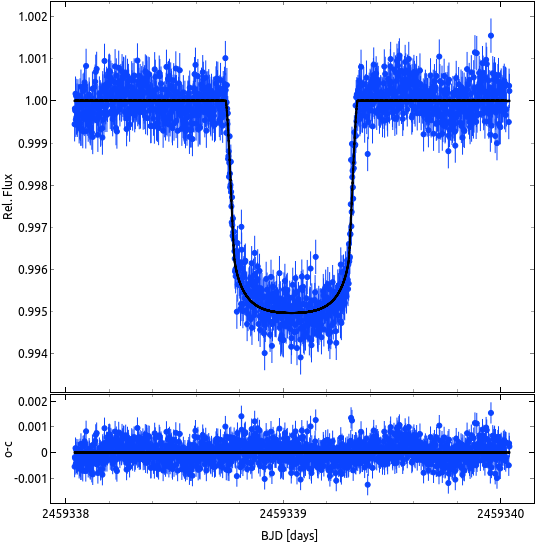}
\centering
\caption{\label{fig:TESS}
Baseline-corrected sector\,38 {\it TESS} data (relative flux) for HD\,114082 around the transit event with the best-fit model of the transit overlaid (black line). The bottom panel shows the residuals of the fit.
}
\end{figure}


Figs.\,\ref{fig:RV} to \ref{fig:TESS} show the results. 
The RVs derived from the FEROS and HARPS spectra are shown in Fig.\,\ref{fig:RV} together with the best-fit model. Fig.\,\ref{fig:phase_pl1} shows the phase-folded data and model, and Fig.\,\ref{fig:TESS} shows part of the TESS data around the transit event. The vertical blue line in Figs.\,\ref{fig:RV} and \ref{fig:phase_pl1} indicates the mid-transit time. 

From the best-fit model we derived a companion mass $M_{\rm pl} = 8.0\pm1.0\,{\rm M_{\rm Jup}}$, a radius $R_{\rm pl} = 1.0\pm0.03\,{\rm R_{\rm Jup}}$, a semi-major axis $a$ = 0.51$\pm$0.01\,au, and an orbital period $P$ = 109.75$^{+0.37}_{-0.40}$\,d{\footnote{This orbital period implies that the first two transit events in 2023 will occur on 22 February (BJD = 2459997.5$^{+2.4}_{-2.2}$\,d) and on 11$^{\rm }$\  June (BJD = 2460107.3$^{+2.8}_{-2.6}$\,d). This is before and after HD\,114082 will be observed with TESS again in sectors 64 and 65 6 April \,--\, 2 June, 2023).}}. All key physical parameters are listed in Table\,\ref{tab:plan_pars}. The MLP periodogram of the residuals from the best-fit model (see Fig.\,\ref{fig:RVmlp_oc}) shows that all the signals of unknown origin are no longer significant. The reason probably is that the best-fit model is much more complex than the sinusoidal fit used to compute the initial MLP power spectrum of Fig.\,\ref{fig:RVgls}.


 \begin{table}
 \caption{Basic planetary parameters of HD\,114082\,b.}
\label{tab:plan_pars}
\centering          
\begin{tabular}{llll}
\hline\hline
Parameter & Unit & Value & Error \\
\hline
$M_{{\rm pl}}$     & [M$_{\rm{Jup}}$]       & 8.0 & $\pm$1.0  \\ 
$R_{{\rm pl}}$     & [R$_{\rm{Jup}}$]       & 1.00 & $\pm$0.03    \\ 
$K$                & [m\,s$^{-1}$]          & 287 & --38/+35   \\
$e$                &   \ldots               & 0.395 & --0.039/+0.037    \\
$a$                & [au]                   & 0.5109 & (--8.3/+8.1) $\times 10^{-3}$        \\
$P$                & [d]                    & 109.75  & --0.37/+0.40 \\
$\omega_b$         & [deg]                  & 211.7 & --3.5/+3.6     \\
$i$                & [deg]                  & 90.00 & --0.18/+0.15   \\
$t_{\circ}$--BJD2450000             & [d]              &  9339.03854 & (--4.9/+5.3) $\times 10^{-4}$  \\  
$a$/R$_\star$      &   \ldots               & 70.4 & --1.8/+1.7   \\ 
$R$/R$_\star$      &   \ldots               & 0.06718 & (--3.5/+3.3) $\times 10^{-4}$     \\ 
RV$_{{\rm off,FEROS}}$ & [m\,s$^{-1}$]      & 10411 & $\pm$21       \\ 
RV$_{{\rm off,HARPS}}$ & [m\,s$^{-1}$]      & -267  & --93/+97       \\ 
RV$_{{\rm jitt,FEROS}}$ & [m\,s$^{-1}$]     & 146   & --18/+21        \\ 
RV$_{{\rm jitt,HARPS}}$ & [m\,s$^{-1}$]     & 283   & --48/+42    \\
Transit$_{{\rm off,TESS}}$ & [ppm]   & 26      & --11/+12                \\
Transit$_{{\rm jitt,TESS}}$ & [ppm]  & 185.56     & $\pm$15               \\ 
 $u_1$ &   \ldots          & 0.089   & --0.056/+0.065           \\
 $u_2$ &   \ldots          & 0.46    & --0.12/+0.11           \\
\hline 
\end{tabular}

\end{table}
 
 \subsection{Stellar activity analysis}
 \label{ssec:res:activity}

We derived the activity indicators for all spectra available for our analysis using the tools described in Section\,\ref{sec:Obs}. We analyzed the generalized Lomb–Scargle periodograms \citep[GLS;][]{Zechmeister2009} of the FEROS and HARPS activity indicators looking for correlations between these indicators and the RVs of the corresponding instrument (see Figs.\,\ref{fig:activity_harps} and \ref{fig:activity_feros} and Table\,\ref{tab:activityCorelations} in Appendix\,\ref{sec:activity} for more details). Only two of all the activity indicators are significantly correlated with the RVs: the BS measured by FEROS, and the CRX measured by HARPS. These correlations have Pearson's correlation coefficients and two-tailed p-values of 0.69, 4.21$\times 10^{-10}$ and 0.81, 5.48$\times 10^{-5}$, respectively. These correlations may indicate that the linear trend, for which we have strong evidence in the RV data (see Section\,\ref{ssec:app:mod}), is most probably related to long-term activity, perhaps an activity cycle, but not to the presence of a larger companion on an outer orbit. 
One way to create a positive BS(RV) correlation would be blending by the line system of a sufficiently bright companion, which is not supported by our mass determination.
We exclude that the RV signal at the 109.8~d period is related to activity because the activity indices from FEROS and HARPS do not show any significant signals near this period (see Figs.\,\ref{fig:activity_harps} and \ref{fig:activity_feros}). We also conclude that the 109.8\,d period cannot be related to stellar rotation, which we estimated to be $P_{{\rm rot}}\leq1.924$\,d (see Table\,\ref{tab:phys_pars}). 
 
  \section{Discussion}
 \label{sec:discussion}

\subsection{Uncertainties introduced by activity}
 \label{ssec:dis:activity}

Except for stellar activity, which manifests itself in significant correlations between the FEROS BS and RVs as well as the HARPS CRX and RVs and a linear trend that we associate with the long-term activity, no other strong activity effects are seen in more than one single indicator. Therefore, the HARPS and FEROS spectra do not support the presence of pronounced activity on HD\,114082.
Early-F dwarfs have shallow convection zones and are therefore not very active in general. However, the considerably large $v\sin(i)$\ of this star may support a certain level of activity. 

HD\,114082 is a rapidly rotating F star ($v\sin(i)$\ = 39.2$\pm$0.5\,km\,s$^{-1}$) that has far fever lines than later-type stars. It also shows substantial rotational broadening. Therefore, the instrumental RV precision is rather poor. Additionally, the star exhibits a large RV scatter, which we modeled as RV jitter quadratically added to the error budget in our fitting. The RV jitter amplitudes determined for both the FEROS and HARPS data individually are  high, 146\,m\,s$^{-1}$ and 283\,m\,s$^{-1}$, respectively. 
This excess uncertainty may also indicate some short-term variability, such as rapid pulsations that are not well-sampled by our data. The star is located near the edge of the instability strip of the Hertzsprung–Russell diagram. Because of its young age, it has not yet fully arrived on the main sequence. 
This places it in the range of $\delta $\,Scuti variables \citep{Breger1972,Breger2000,Murphy2015}, which display multiperiodic $p-$ and $g$-mode pulsations with typical periods of the fundamental mode of a few hours (and more rapid pulsations of some of the harmonics). Although no such periodic pulsations are immediately evident in the GLS periodogram of the TESS data down to periods of $\sim$5\,min, they might still be present but be much weaker than in HD\,115820, for instance, where clearly visible $\delta$-Scuti pulsations between 42 and 68\,min cause a large (6.3\,km/s) RV jitter \citep{zakhozhay2022}. This is well in line with the report by \citet{Grandjean2021} and the analysis of \citet{lagrange2009}, who described that the RV jitter in young A--F5 stars is dominated by pulsations.


\subsection{Astrometry}
 \label{ssec:dis:astrom}

Hipparcos \citep{hipparcos,leeuwen2010} lists an uncritical goodness-of-fit parameter and no multiplicity flag for HD\,114082. 
Gaia \citep[DR3,][]{gaia_mission,GDR3} lists a renormalized unit weight error (RUWE) of 0.94, which is fully consistent with the hypothesis that none of the individual measurements deviates significantly from the single-body solution and with an astrometric excess noise of 149\,$\,\upmu$arcsec. While the value of the latter is considered uncritical for the validity of the single-body astrometric solution, it is significantly higher than the expected maximum astrometric signal of HD\,114082 due to its companion ($\sim$25$\,\upmu$arcsec), which thus remains completely hidden in the excess noise. Furthermore, no significant proper motion anomaly is detected between the long-term HIP-Gaia proper motion and the Gaia-only proper motion vector (PMaG2) according to the formalism described by \citet{kervella2019}, most likely also owing to the smallness and short period of the expected astrometric amplitude. Hence, the available astrometric data do not reveal the existence of this companion, nor do they contradict its existence.



\subsection[HD\,114082\,b in mass-radius space]{HD\,114082\,b in mass--radius space}
 \label{ssec:dis:rpl-mpl}

\begin{figure}[!t]
\includegraphics[width=0.48\textwidth]{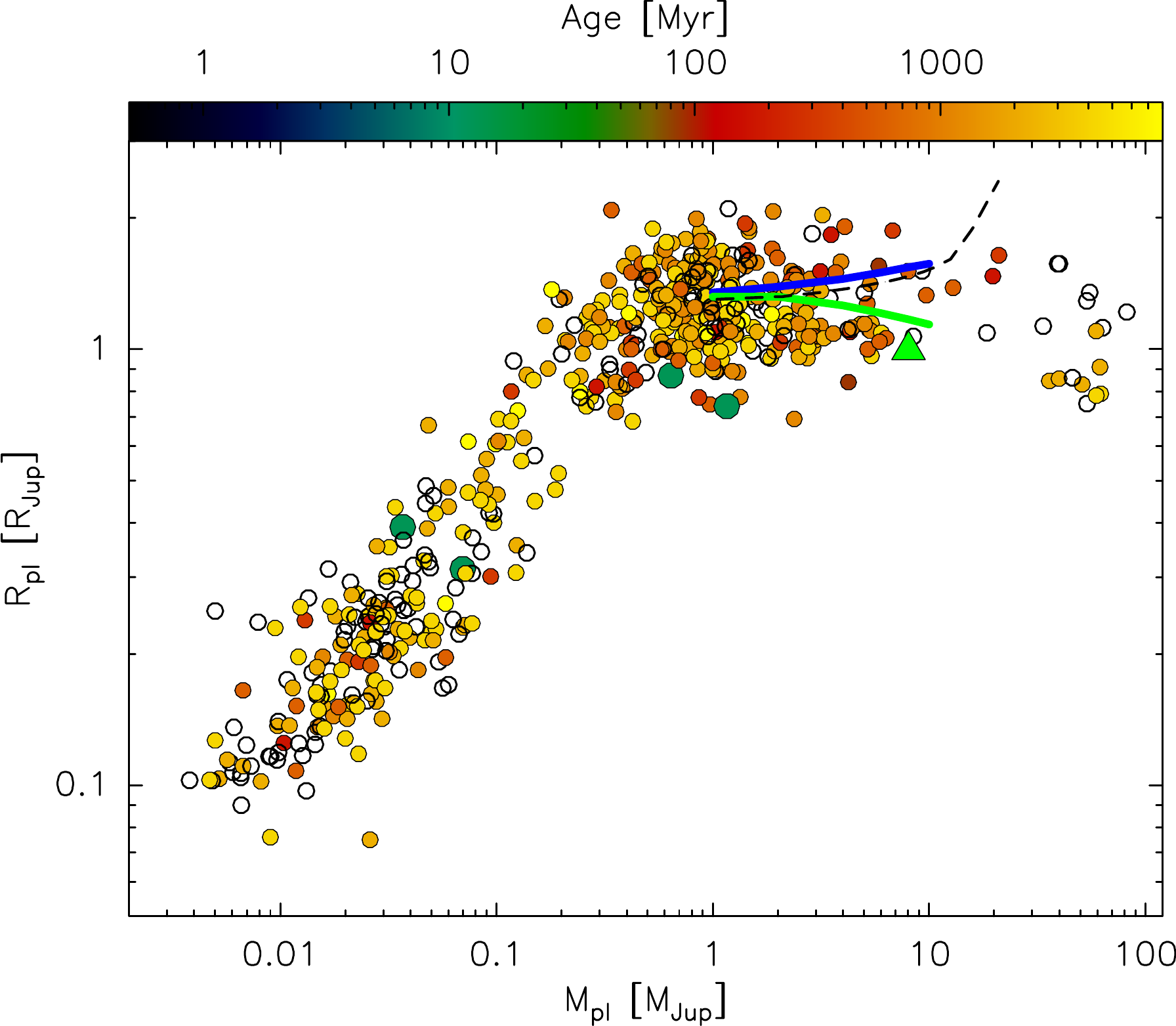}
\centering
\caption{\label{fig:Rpl_Mpl}
Mass--radius distribution of known exoplanets (and brown dwarfs) with radii derived from primary transits and masses derived from RVs (562 planets and companions, excluding upper mass limits). Stellar ages are taken from \url{Exoplanet.eu} and are color-coded. Stars with ages $<$50\,Myr (all green) are shown as larger circles. Stars without known ages are marked as empty circles. The bright green triangle denotes HD\,114082\,b. Error bars, where known, are smaller than the symbol sizes. Thick colored lines mark isochrones for ages of 10-20\,Myr from \citet{fortney2008} for hot-start models (blue) and cold-start models (green). The dashed black line shows the isochrone from the combined COND \citep{baraffe2003} and DUSTY \citep{chabrier2000} models.
}
\end{figure}

Fig.\,\ref{fig:Rpl_Mpl} shows the distribution of all 562 known companions to stars with radii derived from primary transits and masses derived from RVs, excluding planets with only upper-limit mass estimates, together with HD\,114082\,b and color-coded ages. Of these, 408 planets have host stars with ages listed on \url{exoplanet.eu}.
While all values are initially retrieved from \url{Exoplanet.eu}, ages, masses, and radii of the few systems that are younger than 100\,Myr were verified from the respective latest original papers and the values from these papers were used. 
HD\,114082\,b, with 15$\pm$6\,Myr, might be the youngest, and one of only five known exoplanets with ages younger than 30\,Myr for which we know both mass and radius. The four other planets are V\,1298\,Tau\,b and e with 20$\pm$10\,Myr \citep{SuarezMascareno2022} and AU\,Mic\,b and c with 22$\pm$3\,Myr \citep{plavchan2020,martioli2021,zicher2022}. The next-youngest system with known mass and radius is CoroT-20\,b with 100$\pm$40\,Myr \citep{raetz2019}.

Fig.\,\ref{fig:Rpl_Mpl} shows that the measured radius of HD\,114082\,b is close to what cold-start evolutionary models \citep{fortney2008} predict for a 7--10\,M$_{\rm Jup}$ mass planet at age 10--20\,Myr. It is less consistent with the prediction of hot-start models, both from \citet{fortney2008} and from the combined COND \citep{baraffe2003} and DUSTY \citep{chabrier2000} models. 
Together with the two other young ($<$30\,Myr) giant planets (GPs) with known mass and radius (V\,1298\,Tau\,b and e; the two AU\,Mic planets have a lower mass and are rocky and thus probe different physics), we have now three young GPs with measured radii that are significantly smaller than what is predicted by most commonly adopted hot-start models, and not yet a single young GP that agrees with these models. 

\subsection{Comparison with similar planetary systems}

HD\,114082\,b is not the only massive planet found orbiting an F star.
Similar systems include 30\,Ari\,B\,b, a massive planet in a hierarchical stellar triple system \citep{guenther09,kane15}, and the single-star planets HD\,23596\,b \citep{perrier03,wittenmyer09}, HD\,89744\,b \citep{korzennik00,valenti05,wittenmyer19}, HD\,86264\,b \citep{fischer09}, and HD\,1666\,b \citep{harakawa15}. 
HD\,114082 is much younger, and its companion also has the shortest orbital period of the other similar planets mentioned, which range from about 250\,days to just over four years. Otherwise, these systems are similar, however.

Furthermore, a similar population of planets has been found orbiting stars in a more evolved evolutionary stage, mostly K~giants. Almost one-third of all planets and brown dwarfs found around G and K~giants (44 out of 141\footnote{See \url{https://www.lsw.uni-heidelberg.de/users/sreffert/giantplanets/}.}) have masses in excess of 6\,M$_{\mbox{\tiny Jup}}$, and the period distribution is also similar, ranging from about 100 days to 11 years, with a maximum of about 500\,days. 

HD\,114082 is again at the lower end of the period range, but is otherwise similar to the planets around more evolved stars with comparable masses. The most similar system to HD\,114082 with an evolved host star is TYC\,4282-605-1 \citep{gonzalez17}, which hosts a 10\,M$_{\mbox{\tiny Jup}}$ planet on a mildly eccentric 102-day orbit. 
We conclude that the planet orbiting HD\,114082 is not an exception, although no system with such a young age was known so far.
%


 \section{Summary}
 \label{summary}
We presented the results of four years of observations of HD\,114082 with the FEROS spectrograph. The combined analysis of our own FEROS data together with the archival HARPS spectra and TESS photometric data indicates the presence of a GP with a mass of $M_2 = 8.0\pm1.0\,{\rm M_{Jup}}$ on a 109.8$\pm$0.4\,d orbit with $a$ = 0.51$\pm$0.01\,au and $e$ = 0.4$\pm$0.04. The orbit of HD\,114082\,b is thus comparable to the Mercury orbit in our own Solar System, but it is more eccentric.
Based on one transit event in the TESS photometry, we were able to constrain the planetary radius to $1.0\pm0.03\,{\rm R_{Jup}}$, which compares well to the very few other exoplanets in this mass range with known masses and radii. 

HD\,114082\,b might be the youngest such massive GP for which both its mass and radius have been determined observationally. It is also one out of only three young GPs in the 10--30\,Myr age range for which both its mass and radius are known.
It is therefore probably the first real model-constraining GP in mass--radius space based on which we can distinguish between hot- and cold-start models. It is significantly more consistent with the latter.
Together with the two other young ($<$30\,Myr) GPs with known mass and radius (V\,1298\,Tau\,b and e), we have now three young giant planets with measured radii that are significantly smaller than what is predicted by most of the commonly adopted hot-start models. 
Although certainly more such young and massive objects with both observationally constrained mass and radius are needed, this accumulating observational evidence should start triggering an adaption of the evolutionary models.
HD\,114082\,b is also one of the rare cases in which a planetary companion to a star was first discovered in RVs and was then confirmed through a TESS transit.


\begin{acknowledgements}
The authors thank Coryn Bailer-Jones, Christoph Mordasini and 
Remo Burn for helpful discussions and feedback. 
O.Z.\ acknowledges support  within the framework of the Ukraine aid package for individual grants of the Max-Planck Society 2022.
T.T.\ acknowledges support by the BNSF program ``VIHREN-2021'' project No.~KP-06-DV/5.
Th.H.\ acknowledges support from the European Research Council under the European Union’s Horizon 2020 research and innovation program under grant agreement No.~832428-Origins.
S.R.\ and G.-D.M.\ acknowledge support of the DFG (German Science Foundation) priority program SPP~1992 ``Exploring the Diversity of Extrasolar Planets'' (RE~2694/7-1 and MA~9185/1-1).
R.B.\ acknowledges support from FONDECYT Project 11200751 and from project IC120009 ``Millennium Institute of Astrophysics (MAS)'' of the Millenium Science Initiative.
J.I.V.\ acknowledges support of CONICYT-PFCHA/Doctorado Nacional-21191829.
This research made use of the SIMBAD database, operated at the CDS, Strasbourg, France.
G.-D.M. also acknowledges the support from the Swiss National Science Foundation under grant 200021\_204847 ``PlanetsInTime''.
This work presents results from the European Space Agency (ESA) space mission Gaia. Gaia data are being processed by the Gaia Data Processing and Analysis Consortium (DPAC). Funding for the DPAC is provided by national institutions, in particular the institutions participating in the Gaia MultiLateral Agreement (MLA).
Parts of this work have been carried out within the framework of the NCCR PlanetS supported by the Swiss National Science Foundation.
This research has made use of data obtained from or tools provided by the portal \url{exoplanet.eu} of The Extrasolar Planets Encyclopaedia. This work made use of \texttt{tpfplotter} by J. Lillo-Box (publicly available in www.github.com/jlillo/tpfplotter), which also made use of the python packages \texttt{astropy}, \texttt{lightkurve}, \texttt{matplotlib} and \texttt{numpy}.
We also wish to thank the anonymous referee for constructive criticism that helped to improve the clarity of the paper.
\end{acknowledgements}


\bibliographystyle{aa}
\bibliography{zkhbib}


\begin{appendix}

\section{Supplementary material}
\label{sec:supplement}


\subsection{Target pixel image of HD\,11408 in TESS sector 38}
\label{sec:TPF_S38}

\begin{figure}[htb]
\includegraphics[width=0.47\textwidth]{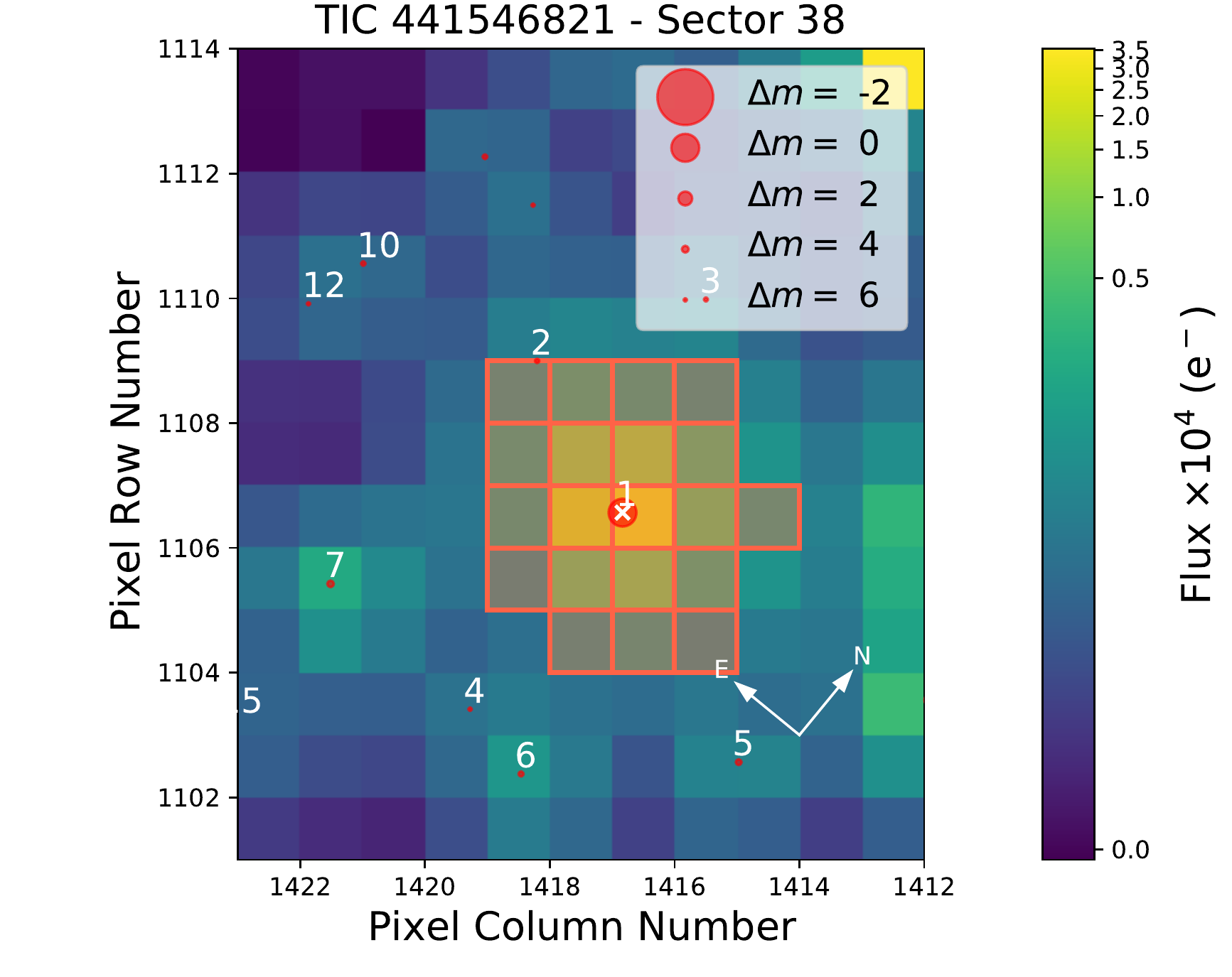}
\centering
\caption{\label{fig:TPF_S38}
Target pixel file image of HD\,114082 in TESS sector 38.
}
\end{figure}

Fig.\,\ref{fig:TPF_S38} shows the target pixel file (TPF) image of HD\,114082 in TESS sector 38. There is one faint object at the edge of the TESS aperture (number 2 in Fig.\,\ref{fig:TPF_S38}) that is very faint (Gmag\,=\,12.56, as compared to 8.1 for HD\,114082) and causes a dilution of 0.984 to the transit light curve of HD\,114082. This contamination dilution was already corrected for in the pre-search data conditioning simple aperture photometry TESS data \citep[PDC-SAP;][see Sect.\,\ref{ssec:app:mod}]{Smith2012, Stumpe12} we used.

 \subsection{Companion signature in TESS data}
 \label{ssec:app:anasep}

The only useful photometric TESS time-series data available (sector 38) show one single transit, which very likely indicates the presence of a bound companion with an orbital period $>27$\,d (the length of the TESS observing sector). The relative transit depth of  $\approx0.5$\% translates into a companion radius of $\approx$1\,R$_{\rm Jup}$, which is indicative of a companion mass between $0.3\,{\rm M}_{\rm Jup}$\ and the brown dwarf mass regime (Fig.\,\ref{fig:Rpl_Mpl}). The transit duration of $\approx0.53\pm0.01$\,d translates into an upper limit to the instantaneous orbital speed (because of the unknown impact parameter) of $\lesssim$44\,km\,s$^{-1}$. Assuming ${\rm M}_{\rm Pl}\ll{\rm M}_{\ast}$ and allowing for a moderate eccentricity of up to $e=0.2$, Kepler's third law implies a minimum orbital period of $P^{\prime}_{\rm min}\approx 190_{-140}^{+100}$\,d.
Since both the orbital period range and the mass range of the transiting companion overlap with the values inferred directly from the RV data alone, we can reasonably assume that both signals are caused by the same companion. We therefore attempted a combined analysis of both data sets in Sect.\,\ref{ssec:res:anacomb}. 


\subsection{Modeling procedure}
\label{ssec:app:mod}

For a combined fit of RVs (the reduced data are listed in Tables~\ref{table:HARPS_RVs} and \ref{table:FEROS_RVs}) and TESS photometry, we used the \texttt{Exo-Striker} toolbox, which allowed us to simultaneously fit both types of data. We used the publically available PDC-SAP TESS data \citep{Smith2012, Stumpe12}, which were corrected for contamination from nearby stars and instrumental systematics. We further detrended and normalized the PDC-SAP light curves with a squared exponential Gaussian processing kernel using the \texttt{Wotan~Python} package~\citep{Hippke2019}, which is integrated into  \texttt{Exo-Striker}. Some obvious photometry outliers were also removed, and the data were cut to $\sim$ $\pm$1~d from the mid-transit to save computational time because the rest of the out-of-transit data do not contain useful information. The TESS data (Fig.\,\ref{fig:TESS}) seem to show an asymmetry during the transit phase, which leads to a slightly inclined bottom of the light curve. So far, it is not clear whether this is a genuine stellar effect or if it arises because no detrending was applied to the transit phase. 

Orbital posteriors were obtained with the \texttt{Nested Sampling} (NS) technique \citep{Skilling2004} using the {\tt dynesty} sampler \citep{Speagle2020}. Our parameter priors are listed in Table\,\ref{table:priors}. Our NS setup consists of 50 live points per fitted parameter, which we sampled via a random walk focused on converging on the posterior probability distribution. We computed the Bayesian log-evidence $\ln(Z)$ to qualitatively compare a Keplerian fit to a Keplerian-and-linear-trend fit.

Our results show that the Bayesian evidence of the simpler model without a linear trend on the RVs has $\ln Z=8686.25\pm 0.20$, whereas the Keplerian model with trend has $\ln Z= 8711.54\pm 0.23$. The inclusion of the linear trend leads to a significant improvement over the trend-free model \citep[i.e., $\Delta \ln Z > 25$, see,][]{Trotta2008}, which additionally justifies use of the model with the linear trend. 
 
\begin{figure}[!t]  
\includegraphics[width=0.48\textwidth]{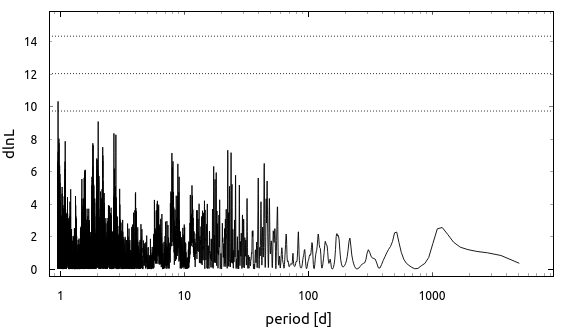}
\centering
\caption{\label{fig:RVmlp_oc}
MLP power of residuals (bottom panel of Fig.\,\ref{fig:RV}). The horizontal lines reflect the 0.1\%, 1\%, and 10\% false-alarm probabilities (from top to bottom).
}
\end{figure}

The weighted rms from the best fit are 158.8 and 290.2~m~s$^{-1}$ for FEROS and HARPS RVs, respectively. The MLP periodogram of the residuals (Fig.~\ref{fig:RVmlp_oc}) shows that all the signals of unknown origin are no longer significant. This is most probably explained by the fact that the best-fit Keplerian model is much more complex than the sinusoidal fit we used to compute the MLP power spectrum. 

\begin{table}[ht]
\centering
\caption{{Fitting parameter priors.  The adopted priors are  $\mathcal{U}$ for uniform and $\mathcal{N}$ for Gaussian.}}
\label{table:priors}
\begin{tabular}{lrrrrrrrr}     
\hline\hline  \noalign{\vskip 0.7mm}
Parameter \hspace{0.0 mm}& HD\,114082\,b \\
\hline \noalign{\vskip 0.7mm}

    $K$ [m\,s$^{-1}$]             & $\mathcal{U}$(    100.00,    700.00)\\
    $P$ [day]                     & $\mathcal{U}$(     50.00,    220.00)\\
    $e$                           & $\mathcal{U}$(      0.00,      0.70)\\
    $\omega$ [deg]                & $\mathcal{U}$(    -90.00,    360.00)\\
    $M_{\rm 0}$ [deg]             & $\mathcal{U}$(      0.00,    360.00)\\
    $i$                           & $\mathcal{N}$(     90.00,     0.20)\\
    $t_{\rm 0}$ [day]             & $\mathcal{U}$(      2459339.00,      2459339.06)\\
    $R_p$/R$_{\star}$                  & $\mathcal{U}$(      0.04,      0.10)\\
    $a$/R$_{\star}$                   & $\mathcal{N}$(     70.00,     2.00)\\
    RV lin. trend [m\,s$^{-1}$\,day$^{-1}$]& $\mathcal{U}$(     -1.00,      0.00)\\
    RV$_{\rm off,FEROS}$ [m\,s$^{-1}$]& $\mathcal{U}$(   9500.00,  11000.00)\\
    RV$_{\rm off,HARPS}$ [m\,s$^{-1}$]& $\mathcal{U}$(   -600.00,    200.00)\\
    RV$_{\rm jitt,FEROS}$ [m\,s$^{-1}$]& $\mathcal{U}$(     50.00,    350.00)\\
    RV$_{\rm jitt,HARPS}$ [m\,s$^{-1}$]& $\mathcal{U}$(     50.00,    350.00)\\
    Transit$_{\rm off}$ [m\,s$^{-1}$]& $\mathcal{U}$(     -100.00,      100.00)\\
    Transit$_{\rm jitt}$ [m\,s$^{-1}$]& $\mathcal{U}$(      10.00,      500.00)\\
    $u_1$& $\mathcal{U}$(     0.00,      1.00)\\
    $u_2$& $\mathcal{U}$(      0.00,      1.00)\\
\hline \noalign{\vskip 0.7mm}
\end{tabular}
\end{table}



\subsection{Activity analysis}
\label{sec:activity}

\begin{figure*}[htb]
\includegraphics[width=0.78\textwidth]{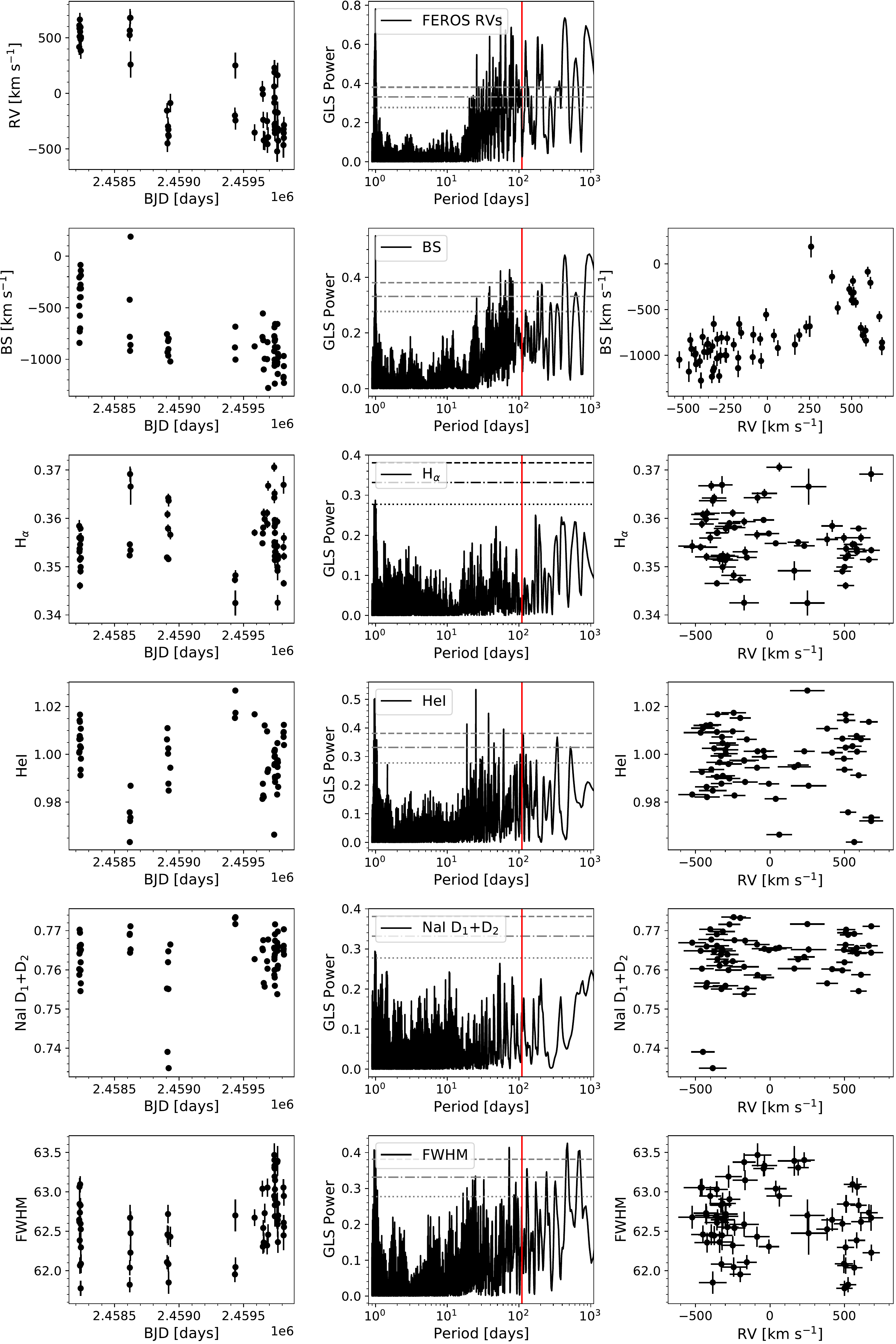}
\centering
\caption{\label{fig:activity_feros}
Time series and GLS of the RVs and activity indicators derived from the FEROS spectra (left and middle panels, respectively). The right panels show the correlation with respect to the FEROS RVs for the derived key activity indicators. The type of activity indicator is given in each of the panels of the middle column. The vertical red line shows the suspected planetary period ($P$ = 109.8\,d), from the model with linear trend of the combined RVs (FEROS+HARPS) and photometric (TESS) data.
}
\end{figure*}

\begin{figure*}[htb]
\includegraphics[width=0.78\textwidth]{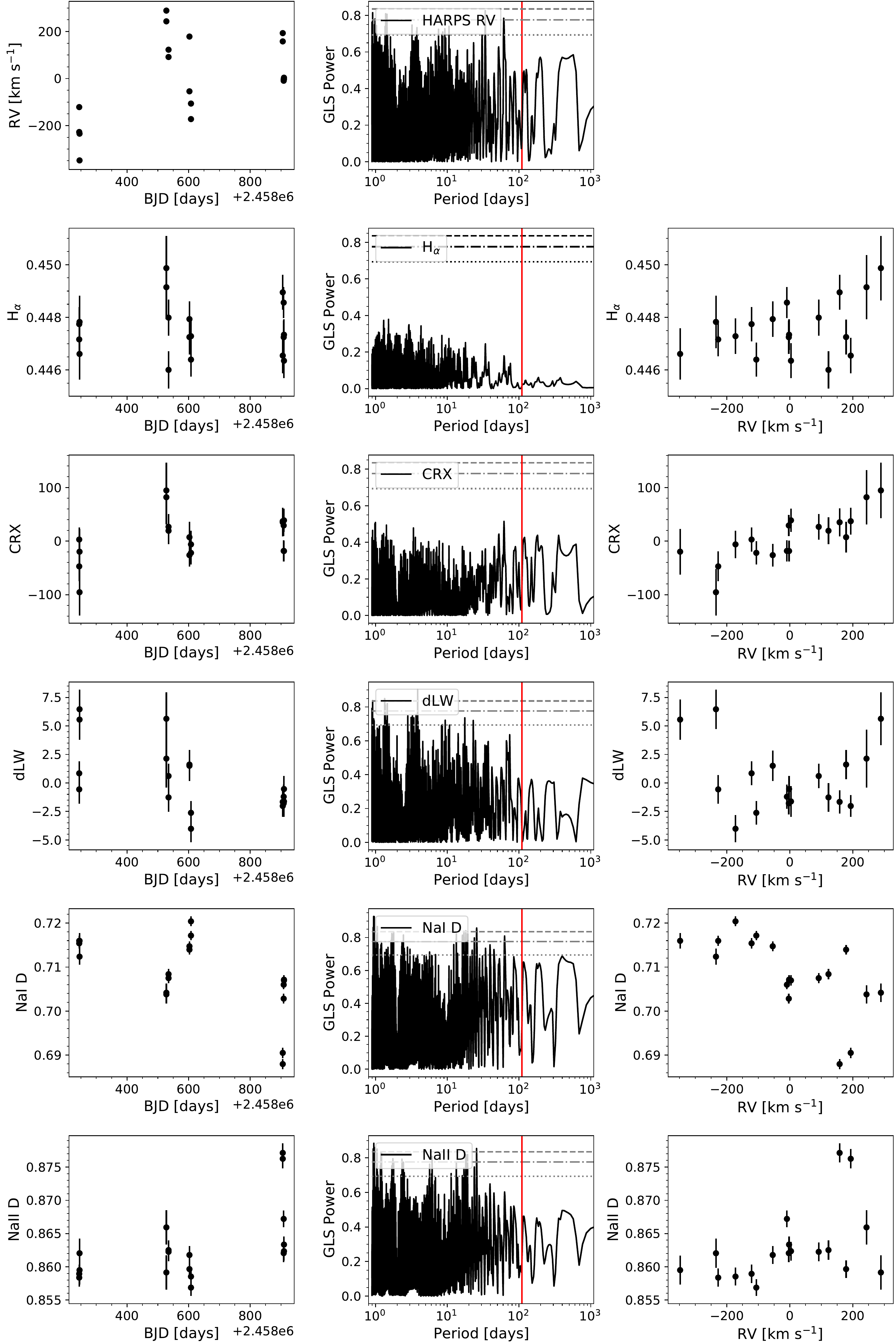}
\centering
\caption{\label{fig:activity_harps}
Same as in Fig.\,\ref{fig:activity_feros}, but for RVs and activity indicators computed from the HARPS spectra.
}
\end{figure*}

Figs.\,\ref{fig:activity_feros} and \ref{fig:activity_harps} show the time series of the activity indicators that we analyzed in this work (for FEROS and HARPS instruments, respectively) in the left columns and their GLS periodograms in the middle panels. The correlations between the activity indicators and the RVs measured by corresponding instrument are shown in the left columns and are listed in Table\,\ref{tab:activityCorelations}. More information about these parameters is presented in Sect.~\ref{sec:Obs}.

\begin{table}
\caption{Pearson's correlation coefficients and two-tailed p-values for correlations of the activity indicators with the respective RVs taken with the same instrument (FEROS or HARPS).}
\label{tab:activityCorelations}
\centering          
\begin{tabular}{llll}
\hline\hline
Activity indicator & Instrument &r & p\\
\hline
BS               & FEROS       &  0.69 & 4.21$\times 10^{-10}$  \\ 
H$_{\alpha}$     & FEROS       &  -0.14 & 0.28  \\ 
He~I             & FEROS       &  -0.08 & 0.54 \\
Na~I~D$_1$+D$_2$ & FEROS       &  0.18 & 0.15 \\ 
FWHM             & FEROS       &  -0.11 & 0.38  \\ 
H$_{\alpha}$     & HARPS       &  0.42 & 0.08  \\ 
CRX              & HARPS       & 0.81 &5.48$\times 10^{-5}$  \\ 
dLW              & HARPS       &  -0.09 & 0.72  \\ 
Na~I~D           & HARPS       &  -0.66 & 0.003  \\ 
Na~II~D          & HARPS       &  0.47 & 0.05 \\ 
\hline 
\end{tabular}
\end{table}

\begin{table}
\caption{HARPS Doppler measurements of HD\,114082.}
\label{table:HARPS_RVs} 
\centering  
\begin{tabular}{c c c } 
\hline\hline    
\noalign{\vskip 0.5mm}
Epoch [JD] & RV [m\,s$^{-1}$] & $\sigma_{RV}$ [m\,s$^{-1}$]   \\  
\hline     
\noalign{\vskip 0.5mm}    
2458244.634   &   -112.7   &    3.4        \\ 
2458244.641   &   -218.7   &    3.4        \\ 
2458245.660   &   -225.8   &    5.9        \\ 
2458245.667   &   -339.5   &    5.7        \\ 
2458527.633   &   297.8   &    7.0        \\ 
2458527.643   &   252.2   &    6.8        \\ 
2458534.693   &   100.5   &    3.3        \\ 
2458534.700   &   131.5   &    3.5        \\ 
2458602.594   &   187.4   &    3.6        \\ 
2458602.602   &   -45.5   &    3.5        \\ 
2458607.574   &   -97.7   &    3.2        \\ 
2458607.581   &   -163.7   &    3.4        \\ 
2458905.605   &   166.8   &    3.6        \\ 
2458905.615   &   202.1   &    3.6        \\ 
2458908.660   &   5.5   &    3.3        \\ 
2458908.668   &   -0.5   &    3.0        \\ 
2458909.607   &   12.4   &    3.6        \\ 
2458909.616   &   6.3   &    3.1        \\ 
\hline           
\end{tabular}
\end{table}

\begin{table}
\caption{FEROS Doppler measurements of HD\,114082.} 
\label{table:FEROS_RVs} 
\centering  
\begin{tabular}{c c c } 
\hline\hline    
\noalign{\vskip 0.5mm}
Epoch [JD] & RV [m\,s$^{-1}$] & $\sigma_{RV}$ [m\,s$^{-1}$]  \\  
\hline     
\noalign{\vskip 0.5mm}    
2458225.778   &   10899.4   &    72.5        \\ 
2458226.772   &   11093.1   &    65.3        \\ 
2458227.767   &   11066.8   &    63.9        \\ 
2458228.782   &   10993.4   &    62.1        \\ 
2458229.787   &   11063.8   &    57.7        \\ 
2458230.741   &   11146.1   &    59.7        \\ 
2458231.745   &   10990.0   &    57.6        \\ 
2458232.844   &   11036.4   &    60.2        \\ 
2458235.527   &   11077.3   &    55.2        \\ 
2458235.708   &   10968.2   &    57.9        \\ 
2458236.765   &   10982.3   &    55.4        \\ 
2458237.838   &   10865.7   &    72.7        \\ 
2458238.667   &   10989.6   &    58.4        \\ 
2458238.805   &   10978.7   &    61.6        \\ 
2458616.737   &   11006.3   &    54.8        \\ 
2458617.735   &   11047.7   &    56.2        \\ 
2458619.800   &   11161.3   &    80.0        \\ 
2458622.590   &   11162.0   &    57.8        \\ 
2458623.681   &   10741.5   &    118.7        \\ 
2458905.872   &   10326.7   &    67.4        \\ 
2458909.785   &   10032.2   &    79.2        \\ 
2458913.836   &   10185.2   &    80.2        \\ 
2458915.802   &   10107.3   &    82.7        \\ 
2458916.817   &   10156.2   &    71.3        \\ 
2458918.723   &   10098.0   &    92.9        \\ 
2458931.840   &   10395.2   &    84.3        \\ 
2459432.503   &   10282.7   &    70.8        \\ 
2459435.508   &   10733.1   &    116.6        \\ 
2459436.495   &   10238.3   &    82.5        \\ 
2459584.827   &   10129.5   &    75.0        \\ 
2459645.760   &   10520.7   &    74.0        \\ 
2459648.828   &   10474.6   &    69.6        \\ 
2459651.733   &   10243.7   &    74.2        \\ 
2459655.692   &   10059.6   &    85.7        \\ 
2459662.775   &   10053.0   &    83.8        \\ 
2459681.742   &   10232.3   &    79.3        \\ 
2459684.756   &   10025.0   &    79.7        \\ 
2459689.733   &   10088.2   &    88.0        \\ 
2459736.494   &   10544.4   &    86.7        \\ 
2459737.505   &   10398.5   &    93.5        \\ 
2459738.496   &   10671.2   &    70.9        \\ 
2459738.590   &   10712.2   &    70.2        \\ 
2459739.466   &   10444.8   &    86.0        \\ 
2459739.717   &   10203.7   &    89.7        \\ 
2459740.469   &   10439.3   &    70.5        \\ 
2459741.466   &   10153.1   &    71.0        \\ 
2459741.704   &   10183.1   &    86.4        \\ 
2459742.459   &   10125.2   &    74.0        \\ 
2459743.490   &   10209.9   &    74.0        \\ 
2459743.683   &   10315.2   &    83.6        \\ 
2459759.580   &   9960.0   &    93.2        \\ 
2459760.619   &   10309.2   &    77.5        \\ 
2459761.568   &   10164.2   &    89.4        \\ 
2459762.501   &   10645.5   &    111.2        \\ 
2459763.518   &   10308.4   &    99.0        \\ 
2459764.523   &   10165.6   &    81.3        \\ 
2459765.481   &   10143.8   &    88.9        \\ 
2459767.478   &   10056.4   &    78.6        \\ 
2459808.470   &   10196.4   &    74.8        \\ 
2459809.481   &   10016.6   &    113.4        \\ 
2459810.497   &   10160.0   &    70.6        \\ 
2459811.506   &   10125.8   &    82.0        \\ 
2459812.472   &   10081.3   &    74.0        \\ 
\hline           
\end{tabular}
\end{table}

\end{appendix}

\end{document}